\documentclass[runningheads]{llncs}
\usepackage[T1]{fontenc}
\usepackage{graphicx,verbatim}

\usepackage[
    colorlinks=true,
    linkcolor=red,
    citecolor=green,
    urlcolor=black
]{hyperref}

\usepackage{orcidlink}

\begin{document}
\title{Virtual Dosimetrists: A Radiotherapy Training "Flight Simulator"}
\author{Skylar Gay\inst{1,2}\thanks{Corresponding author: \email{sgay1@mdanderson.org}}\orcidlink{0000-0003-4659-0766} \and
Tucker Netherton \inst{2}\orcidlink{0000-0003-1583-7121} \and
Barbara Marquez\inst{1,2}\orcidlink{0000-0002-0070-6930} \and
Raymond Mumme \inst{2} \orcidlink{0000-0001-7383-6122} \and
Mary Gronberg \inst{3} \orcidlink{0000-0002-0121-9579} \and
Brent Parker \inst{4} \orcidlink{0000-0001-7015-8805} \and
Chelsea Pinnix \inst{2} \orcidlink{0000-0003-3982-3664} \and
Sanjay Shete \inst{2} \orcidlink{0000-0001-7622-376X} \and
Carlos Cardenas \inst{5} \orcidlink{0000-0003-1414-3849} \and
Laurence Court \inst{2} \orcidlink{0000-0002-3241-6145}}

\authorrunning{S. Gay, et al.}

\institute{The University of Texas MD Anderson Cancer Center UTHealth Graduate School of Biomedical Sciences, Houston TX 77030 \and
The University of Texas MD Anderson Cancer Center, Houston TX 77030 \and
The University of Texas Southwestern Medical Center, Dallas TX 75390 \and
The University of Texas Medical Branch at Galveston, Galveston TX 77555 \and
The University of Alabama at Birmingham, Birmingham AL 35294
}

\authorrunning{S. Gay, et al.}

\maketitle
\begin{abstract}
Effective education in radiotherapy plan quality review requires a robust, regularly updated set of examples and the flexibility to demonstrate multiple possible planning approaches and their consequences. However, the current clinic-based paradigm does not support these needs. To address this, we have developed “Virtual Dosimetrist” models that can both generate training examples of suboptimal treatment plans and then allow trainees to improve the plan quality through simple natural language prompts, as if communicating with a dosimetrist. The dose generation and modification process is accurate, rapid, and requires only modest resources. This work is the first to combine dose distribution prediction with natural language processing; providing a robust pipeline for both generating suboptimal training plans and allowing trainees to practice their critical plan review and improvement skills that addresses the challenges of the current clinic-based paradigm.


\end{abstract}

\section{Introduction}

Radiation oncology residents report substantial limitation in their grasp of radiotherapy plan quality and review skills \cite{boydRadiationTreatmentPlan2023}. The current clinic-based training contributes heavily to these limitations. There are substantial gaps in presentation of diverse disease sites [unpublished National Cancer Database analysis, 2021]. Furthermore, residents often observe only a single treatment plan rather than the many lower-quality variations which are possible for any given case. Thus, conceptualizing the range of plans needing improvement can be difficult.

Additionally, due to its manual nature, there are substantial time gaps between identifying how a plan may be improved and the generation of an improved radiotherapy plan by the clinical planners. This creates a second challenge in effective training, as understanding which of myriad potential changes are more beneficial becomes difficult with this delayed feedback. Combined, these challenges may spill into patient care, with the propagation of suboptimal treatment plans through the review process and even into patient treatment being well-documented \cite{talcottBlindedProspectiveStudy2020a,mooreQuantifyingUnnecessaryNormal2015b,mooreExperienceBasedQualityControl2011}.

The first challenge, data availability, might be temporarily addressed through multi-institutional radiotherapy plan curation to generate a large training corpus. However, the differences in planning styles across institutions detracts from the educational benefit. It would also be infeasible in typical planning practice to save and export all early-stage treatment plans, so trainees would still not be able to view many lower-quality plan examples. Finally, inadvertent memorization by trainees would reduce the effectiveness of any static training database.

The second challenge, the gaps between residents identifying potential improvements and viewing the results, is even more difficult to address. Planners in a typical clinic have a heavy workload and cannot rapidly generate new plans for educational purposes. In addition, modifying a treatment plan requires minutes to hours. This presents substantial limitations to effective education.

Recently, deep learning dose prediction has shown strong results across a wide variety of treatment sites and planning styles. In particular, knowledge-based approaches have been widely adopted \cite{nguyenIncorporatingHumanLearned2020,babierOpenKBPOpenaccessKnowledgebased2021,gronbergTechnicalNoteDose2021a,gronbergDeepLearningBased,gronbergDeepLearningBased2023}. However, researchers focus on producing high-quality (e.g., Pareto-optimal) dose distributions for clinical use, not lower-quality examples for education. As well, the models are typically “frozen;” the quality of the distributions they generate cannot be adjusted.

Therefore, to address these challenges, we have developed “Virtual Dosimetrist” dose modification models. These cross-modal, novel models require only patient data – a computed-tomography (CT) scan, a starting radiotherapy dose distribution, and a few structure masks – and a natural language prompt describing what to change in the dose. Then, the models can generate new dose distribution of many different qualities. This enables both the production of many training examples, and essentially limitless, real-time abilities for residents to request improvements and receive a modified dose distribution as if generated by a clinical planner in a treatment planning system (TPS). To the best of our knowledge, this is the first work to combine language and dose prediction architectures and the first to demonstrate direct modification of dose distributions following a language prompt.

\section{Methods}
\subsection{Data Generation, Curation, and Augmentation}
In this work, treatment plans were used to train the virtual dosimetrist models, either clinically-approved (e.g., high quality) plans or ones replanned to intentionally decrease quality. Fifty-three head-and-neck (HN) volumetric-modulated arc therapy (VMAT) treatment plans were curated from our clinical practice. All plans were physician-approved and delivered between 2019-2023. The HN site was selected due to its planning complexity – successful design of a virtual dosimetrist for HN increases confidence in applicability to other, less-complex disease sites.

Suboptimal treatment plans were generated using the clinical TPS to have poor organ at risk (OAR) sparing, which leads to increased patient toxicities and reduced quality of life. To generate these lower-quality plans from clinical data, we developed a novel replanning technique. For each OAR with one or more planning objectives, we identified all structures that substantially contributed to the OAR dose, such as expansions or false structures with high anatomical similarities. We also identified all relevant planning objectives for each OAR, e.g. mean dose, maximum dose, and dose-volume histogram (DVH) objectives.

The clinical plans (denoted $D_0$) were highly optimized, so merely relaxing the planning objectives did not increase OAR dose. Therefore, each OAR and relevant structures were temporarily reset to a “target” type to instruct the TPS optimizer to prioritize achieving minimum dose goals. Then the existing dose information was used to adjust the planning objectives so that the OAR dose would be increased. This process was repeated, starting with the newly-replanned dose, for a total of 5 new plans for each OAR (denoted $D_1, ... ,D_5$). To ensure that the dose typically increased for each iteration, a simple line-search algorithm was implemented to update the planning objectives based upon the change to OAR mean dose in previous replanning iterations. Finally, the planning CT, clinical and all replanned doses, target structures, and OAR structures were exported with the dose grid voxel spacing (3mm$\times$3mm$\times$3mm). To increase the diversity of large dose changes, this process was repeated where change to OAR maximum dose was used to update the planning objectives.

Simple curation was conducted so that the models would be trained on clinically realistic data. Any replanned doses that exceeded 8400cGy, or approximately 120\% of the typical prescription, were removed from the dataset. This resulted in 4072 final dose distributions (Table~\ref{tab1}).

\begin{table}
\centering
\caption{Number of dose distributions per OAR in final dataset.}\label{tab1}
\begin{tabular}{|l|r|}
\hline
OAR (\textbf{L}eft/\textbf{R}ight) &  \# Dose Distributions \\
\hline
Parotid (L) & 420 \\
Parotid (R) & 403 \\
Esophagus & 373 \\
Brainstem & 357 \\
Larynx & 355 \\
Spinal Cord & 348 \\
Cochlea (L) & 320 \\
Cochlea (R) & 310 \\
Lungs & 248 \\
Oral Cavity & 238 \\
Brachial Plexus (R) & 142 \\
Brachial Plexus (L) & 125 \\
Submandibular Gland (L) & 111 \\
Mandible & 102 \\
Submandibular Gland (R) & 100 \\
Lens (L) & 60 \\
Lens (R) & 60 \\
\hline
\end{tabular}
\end{table}

To avoid potential bias, a 80\%/20\% train/test dataset split was conducted on the patient level, with 42 patients (3214 dose distributions) in the training set and a final 11 patients (858 dose distributions) reserved for the unseen test set. To improve model generalizability, training was conducted with a five-fold cross-validation scheme, again with the train/validation split conducted on the patient level.

Finally, during model training, the following data processing and augmentation rules were used:
\begin{itemize}
\item The CT Hounsfield unit values were clipped to (-200, 450), and then normalized to the range (0, 1).
\item The dose maps were normalized to the range (0, 1) by dividing all values by 8400. 
\item The target masks were multiplied by their prescription in centigray (cGy), and then divided by 8400.
\item The OAR mask was isotropically dilated by 1.5cm.
\item There was a 67\% chance of the dose feature/label pair being the clinical dose  and a replanned dose (e.g., $D_0$ and $D_3$), instead of two replanned doses (e.g., $D_2$ and $D_3$).
\item To enable bidirectional training, there was a 50\% chance the model would be instructed to increase dose and a 50\% chance to decrease dose. 
\item The input data was randomly cropped to dimensions (96, 112, 176) in the (depth, y, x) dimension, based upon the dimensions of the training set.
\item There was a 20\% probability for each of the following augmentations schemes to be applied in the specified order:
    \begin{itemize}
    \item Random rotation in the x-y plane up to ±3\textdegree
    \item Random zoom in any plane up to ±10\%
    \item Random flip in the x-y plane
    \end{itemize}
\end{itemize}

\subsection{Architecture}
\begin{figure}
\includegraphics[width=\textwidth]{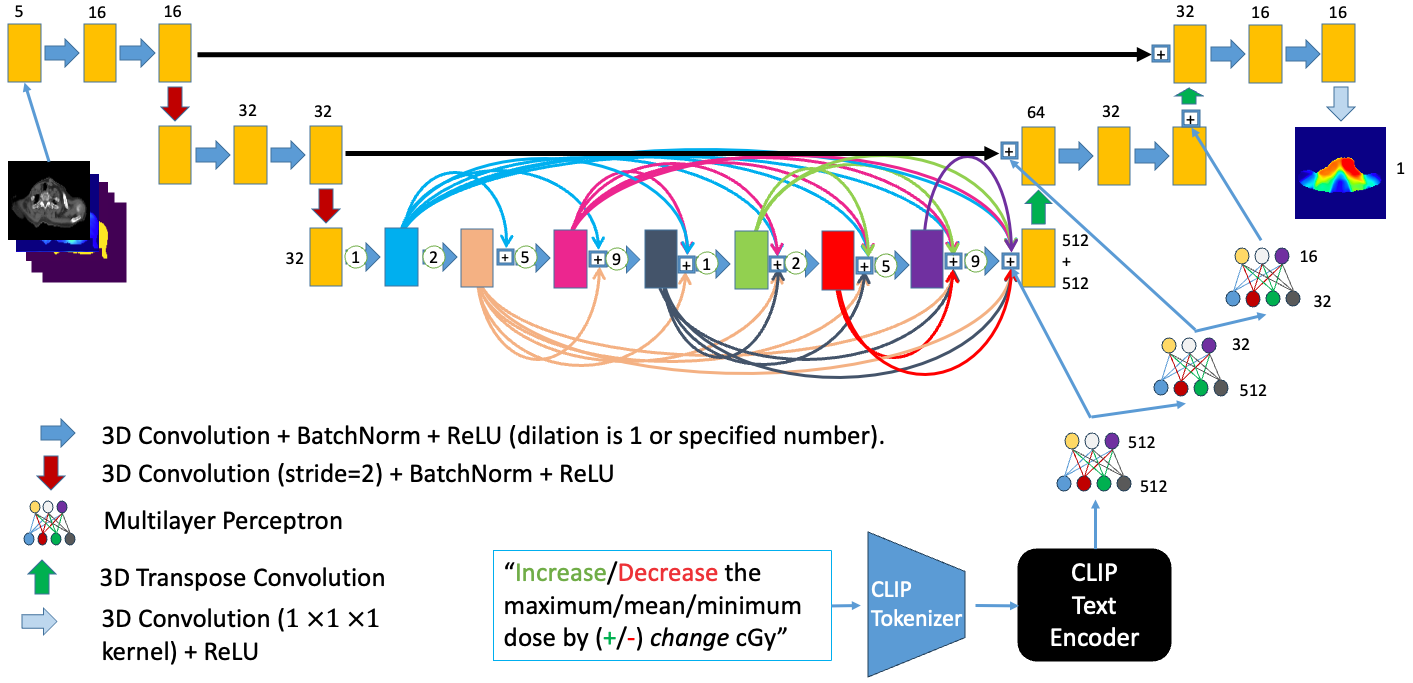}
\caption{Overview of model architecture. Unless otherwise specified, all convolution kernels dimension are 3×3×3, with the number of channels indicated in the figure.} \label{fig1}
\end{figure}

A dual-encoder architecture was designed for this study (Fig.~\ref{fig1}). To process the volumetric data (CT, dose map, potential dose mask, target mask, and dilated OAR mask), the 3D DDU-Net was chosen based upon its strong performance in the dose prediction domain \cite{gronbergTechnicalNoteDose2021a}. To encode the language information describing the dose changes to be made (e.g., "Decrease the mean dose by -400 cGy"), the OpenAI CLIP text encoder was selected based upon its understanding of medical language and reported performance on cross-modal tasks \cite{radfordLearningTransferableVisual2021,liuCLIPDrivenUniversalModel2023,chenDifferenceBERTstyleCLIPstyle2023}. Publicly available pre-trained weights provided by OpenAI were used, with multi-layer perceptrons (MLPs) added for fine-tuning the dose modification tasks. The architecture was written in PyTorch 2.5.1 and trained using a Nvidia A40 GPU. All convolution operations except the final were followed by batch normalization \cite{Ioffe2015} and rectified linear activation unit (ReLU) functions to improve stability and prevent negative outputs. The final convolution was followed by only ReLU to avoid potential shift in output dose values by batch normalization. During train time, batch sizes of 8 were used. The Adam \cite{kingmaAdamMethodStochastic2014} optimizer was selected, with the learning rate set as $8.5\times10^{-4}$, following scaling described in the literature \cite{krizhevskyOneWeirdTrick2014,granziolLearningRatesFunction2022}. If 50 epochs passed without a reduction in the validation data loss, the learning rate was decreased by a factor of 10\%. The models trained for a maximum of 1000 epochs, and terminated early if validation loss did not decrease for 120 epochs.

A weighted mean-squared error loss function focused the model on the region of and surrounding the OAR. Using the dose mask, target mask, and dilated OAR mask, the mean-squared error was calculated for each region. Additional weight was given to the dilated OAR region, as particular accuracy was necessary here – when moving further from the OAR, the output dose should be more similar to the input dose.The final loss function is described in Equation~\ref{eq1}, where dm denotes the dose mask, PTV denotes the target mask, OAR denotes the dilated OAR mask, $D$ denotes the ground truth dose map, and $\hat{D}$ denotes the predicted dose map
\begin{equation}\label{eq1}
\mathcal{L} = 0.2 MSE(D_{dm}, \hat{D}_{dm}) + 0.2 MSE(D_{PTV}, \hat{D}_{PTV}) + 0.6 MSE(D_{OAR}, \hat{D}_{OAR})
\end{equation}

\noindent where MSE is the mean-squared error described by Equation~\ref{eq2}.
\begin{equation}\label{eq2}
MSE(D, \hat{D}) = \frac{1}{n} \sum_{i=1}^n (D_i - \hat{D}_i)^2
\end{equation}

\subsection{Inference and Post Processing Techniques}

As the models were trained on randomly-cropped inputs, a sliding-window approach was adapted from the MONAI project \cite{consortiumMONAIMedicalOpen2024}. Each patch overlapped 50\% with neighboring patches, and Gaussian weighting was used so that predictions from central patches contributed more than predictions from edge patches. This approach was used with each of the five trained models, and the output dose prediction generated as the constant-weighted average of the individual models’ predictions.

To increase precision, a simple loop-based approach was designed in the inference process. Following the initial inference and prompt, the resulting dose difference for the specified OAR was compared to the requested change and metric. If the difference was not less than ±50cGy, a new prompt was generated with the difference and the prediction repeated on the initial inference data. This process was repeated until the inferred dose change was within ±50cGy of the initial prompt, or 10 iterations, whichever occurred first.

\section{Results}
The virtual dosimetrist models were successful at producing new dose distributions with the requested change, both for increasing and decreasing dose to OARs. After selecting an OAR mask, the user-supplied prompt is a simple language input in the format:
\begin{itemize}
    \item Increase the \textit{metric} dose by \textit{amount} cGy
    \newline or
    \item Decrease the \textit{metric} dose by -\textit{amount} cGy
\end{itemize}
where \textit{metric} is “minimum”, “mean”, or “maximum”. On average, the entire pipeline takes only 7 seconds to generate the new dose distributions.

\begin{figure}
\includegraphics[width=\textwidth]{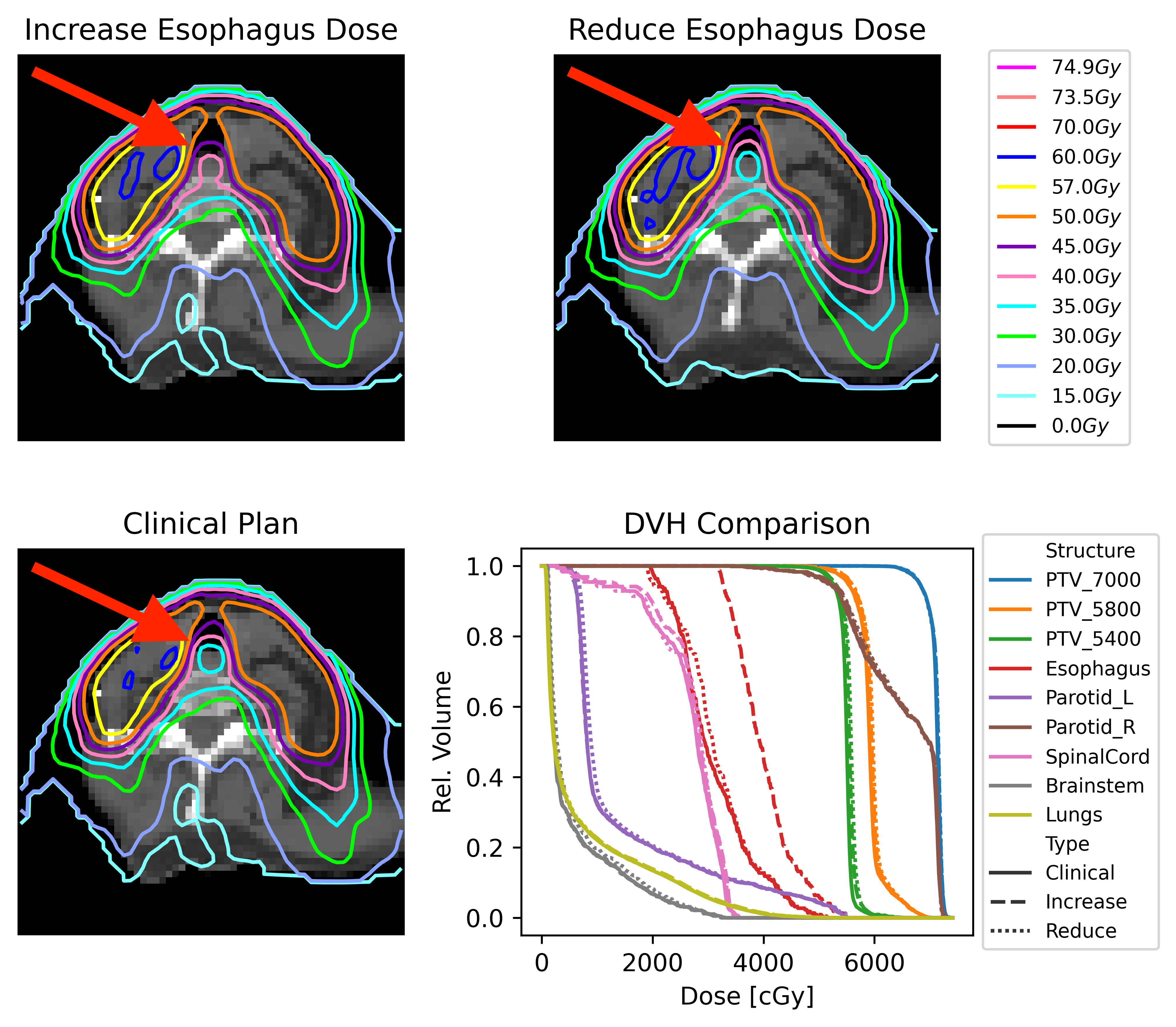}
\caption{A generated lower-quality training example (top left) with reduced esophagus sparing. With the prompt "Decrease the mean dose by -900 cGy", a new dose distribution with improved esophagus sparing was then generated (top right) that is similar to the clinical dose distribution (bottom left). The DVH comparison (bottom right) shows how this tool can both generate lower-quality educational examples and subsequently improve the quality of the examples to be similar to the actual clinical plan. Red arrow: region of the esophagus.} \label{fig2}
\end{figure}

The results for the prompt to increase dose by 500cGy, in order to generate training examples, is shown in Table~\ref{tab2}. The results for prompts to decrease dose by the same amount are similar. Fig.~\ref{fig2} demonstrates the utility of this tool for training. A  training example is first generated by increasing the mean dose to the esophagus. The example quality can then be improved by the trainee through appropriate planning directives, similar to the clinical workflow. This allows trainees to be presented with lower-quality examples and easily practice their plan quality review skills in a clinic-like environment with rapid feedback.

\begin{table}
\centering
\caption{Median change in cGy between predicted and clinical dose to specified dose metric, following the specified prompt to increase dose. $\Delta$Dmin: median change in minimum dose to the OAR. $\Delta$Dmean: median change in mean dose to the OAR: $\Delta$Dmax: median change in maximum dose to the OAR. $\sigma$: standard deviation.}\label{tab2}
Prompt: "Increase the \textit{metric} dose by 500 cGy"
\newline
\begin{tabular}{|l|r|r|r|}
\hline
OAR (\textbf{L}eft/\textbf{R}ight) &  $\Delta$Dmin ($\sigma$) &  $\Delta$Dmean ($\sigma$) &  $\Delta$Dmax ($\sigma$) \\
\hline
Brachial Plexus (L) & 513.6 (61.2) & 557.6 (80.2) & 576.7 (176.6) \\
Brachial Plexus (R) & 585.0 (30.1) & 578.5 (99.8) & 672.9 (92.6) \\
Brainstem & 512.7 (78.3) & 556.1 (54.9) & 644.0 (77.7) \\
Oral Cavity & 559.7 (261.9) & 511.3 (64.6) & 379.8 (232.3) \\
Cochlea (L) & 586.0 (50.0) & 524.8 (29.4) & 507.8 (28.4) \\
Cochlea (R) & 532.5 (53.8) & 500.6 (38.3) & 538.8 (61.9) \\
Esophagus & 344.8 (265.9) & 561.6 (34.1) & 562.2 (66.6) \\
Larynx & 506.6 (253.6) & 646.3 (101.0) & 551.1 (132.5) \\
Lens (L) & 621.7 (19.7) & 548.8 (10.8) & 540.8 (65.6) \\
Lens (R) & 581.0 (28.0) & 503.2 (17.9) & 506.6 (12.7) \\
Lungs & 466.6 (228.8) & 531.7 (38.1) & 612.9 (118.9) \\
Mandible & 564.9 (73.0) & 527.1 (41.1) & 7.0 (14.5) \\
Parotid (L) & 512.2 (76.5) & 619.4 (109.0) & 380.1 (245.5) \\
Parotid (R) & 525.7 (61.7) & 561.9 (92.1) & 414.9 (208.6) \\
Spinal Cord & 511.7 (237.6) & 542.1 (11.6) & 579.0 (94.2) \\
Submandibular Gland (L) & 743.5 (191.2) & 607.8 (37.8) & 386.1 (84.3) \\
Submandibular Gland (R) & 520.9 (212.8) & 549.7 (31.7) & 392.7 (93.0) \\
\hline
\end{tabular}
\end{table}

\section{Discussion}
In this work, we developed techniques to generate new dose distributions through the combination of natural language processing with dose prediction models. The changes are accurate, require only a few seconds for generation, and provide opportunities for training radiation oncology residents and other trainees in plan quality review.

For certain OARs, it may be observed that the magnitude of dose change achievable is less than others. This effect is most strongly pronounced for the oral cavity and mandible, which are near air boundaries and are often less well-defined than other structures in the dataset. There were also somewhat fewer training examples for these structures than many others. Future work could investigate the impact of increasing the example availability for these structures.

To the best of our knowledge, this is the first work to develop a deep learning-based architecture that directly generates and modifies dose distributions following a language prompt. Previous researchers developed “virtual treatment planner” (VTP) models that can adjust planning parameters to improve plan quality and operates in a human-like manner on both in-house \cite{shenOperatingTreatmentPlanning2020b} and commercially-available TPSs \cite{sproutsDevelopmentDeepReinforcement2022,gaoImplementationEvaluationIntelligent2023}, and can interact verbally with a physician \cite{gaoHighDoseRateBrachytherapy2023}. However, the previous work uses deep learning models to control the TPS, rather than adjusting the dose distribution directly based upon the user prompt. It also focuses on improving or adjusting plans based upon physician preference, rather than also being designed to produce lower-quality dose distributions for training. Finally, although it reduces plan generation time from approximately 15 minutes to 3 minutes, our approach is substantially faster at approximately 7 seconds.

\section{Conclusion}
In this work, a novel method to generate dose distributions of varying qualities, and to iteratively improve the qualities, was developed. The changes can be made rapidly, requiring only about 7 seconds on average. This tool shows promise for clinical education applications, particularly in training for plan quality review and improvement, completely independent of any TPS and in an environment that is substantially faster than the current clinic-based approach.

\begin{credits}
\subsubsection{\ackname} The authors acknowledge the support of the High Performance Computing for Research facility at The University of Texas MD Anderson Cancer Center (supported by the NIH/NCI under award number P30CA016672) and the Texas Advanced Computing Center team for providing computational resources that have contributed to the results reported in this manuscript.

\subsubsection{\discintname}
This work was supported by a grant from The University of Texas MD
Anderson Cancer Center, Division of Radiation Oncology. Skylar Gay
received support from the Cancer Answers Scholarship, the Larry Deaven
PhD Fellowship, and the Dr. John J. Kopchick Fellowship. There are no
other conflicts of interest to disclose.
\end{credits}

%
%
%
\newpage
\bibliographystyle{splncs04}
\bibliography{bibliography}

\begin{thebibliography}{10}
\providecommand{\url}[1]{\texttt{#1}}
\providecommand{\urlprefix}{URL }
\providecommand{\doi}[1]{https://doi.org/#1}

\bibitem{babierOpenKBPOpenaccessKnowledgebased2021}
Babier, A., Zhang, B., Mahmood, R., Moore, K.L., Purdie, T.G., McNiven, A.L., Chan, T.C.Y.: {OpenKBP}: {The} open-access knowledge-based planning grand challenge and dataset. Medical Physics  \textbf{48}(9),  5549--5561 (2021). \doi{10.1002/mp.14845}

\bibitem{boydRadiationTreatmentPlan2023}
Boyd, G.H., Vanbenthuysen, L., Jimenez, R.B.: Radiation {Treatment} {Plan} {Evaluation} {Education} in {Residency}: {A} {Needs} {Assessment}. International Journal of Radiation Oncology, Biology, Physics  \textbf{117}(2),  e504--e505 (Oct 2023). \doi{10.1016/j.ijrobp.2023.06.1753}

\bibitem{chenDifferenceBERTstyleCLIPstyle2023}
Chen, Z., Chen, G.H., Diao, S., Wan, X., Wang, B.: On the {Difference} of {BERT}-style and {CLIP}-style {Text} {Encoders} (Jun 2023). \doi{10.48550/arXiv.2306.03678}

\bibitem{gaoHighDoseRateBrachytherapy2023}
Gao, Y., Shen, C., Gonzalez, Y., Jia, X.: High {Dose}-{Rate} {Brachytherapy} {Treatment} {Planning} for {Gynecological} {Cancer} with {Intelligent} and {Conversational} {AI}. AAPM (Jul 2023), \url{https://aapm.confex.com/aapm/2023am/meetingapp.cgi/Paper/6126}

\bibitem{gaoImplementationEvaluationIntelligent2023}
Gao, Y., Shen, C., Jia, X., Kyun~Park, Y.: Implementation and evaluation of an intelligent automatic treatment planning robot for prostate cancer stereotactic body radiation therapy. Radiotherapy and Oncology  \textbf{184},  109685 (Jul 2023). \doi{10.1016/j.radonc.2023.109685}

\bibitem{granziolLearningRatesFunction2022}
Granziol, D., Zohren, S., Roberts, S.: Learning {Rates} as a {Function} of {Batch} {Size}: {A} {Random} {Matrix} {Theory} {Approach} to {Neural} {Network} {Training}. Journal of Machine Learning Research  \textbf{23}(173),  1--65 (2022), \url{http://jmlr.org/papers/v23/20-1258.html}

\bibitem{gronbergDeepLearningBased2023}
Gronberg, M.P., Beadle, B.M., Garden, A.S., Skinner, H., Gay, S., Netherton, T., Cao, W., Cardenas, C.E., Chung, C., Fuentes, D.T., Fuller, C.D., Howell, R.M., Jhingran, A., Lim, T.Y., Marquez, B., Mumme, R., Olanrewaju, A.M., Peterson, C.B., Vazquez, I., Whitaker, T.J., Wooten, Z., Yang, M., Court, L.E.: Deep {Learning}–{Based} {Dose} {Prediction} for {Automated}, {Individualized} {Quality} {Assurance} of {Head} and {Neck} {Radiation} {Therapy} {Plans}. Practical Radiation Oncology  (Jan 2023). \doi{10.1016/j.prro.2022.12.003}

\bibitem{gronbergTechnicalNoteDose2021a}
Gronberg, M.P., Gay, S.S., Netherton, T.J., Rhee, D.J., Court, L.E., Cardenas, C.E.: Technical {Note}: {Dose} prediction for head and neck radiotherapy using a three-dimensional dense dilated {U}-net architecture. Medical Physics  \textbf{48}(9),  5567--5573 (2021). \doi{10.1002/mp.14827}

\bibitem{gronbergDeepLearningBased}
Gronberg, M.P., Jhingran, A., Netherton, T.J., Gay, S.S., Cardenas, C.E., Chung, C., Fuentes, D., Fuller, C.D., Howell, R.M., Khan, M., Lim, T.Y., Marquez, B., Olanrewaju, A.M., Peterson, C.B., Vazquez, I., Whitaker, T.J., Wooten, Z., Yang, M., Court, L.E.: Deep learning–based dose prediction to improve the plan quality of volumetric modulated arc therapy for gynecologic cancers. Medical Physics  \textbf{50}(11),  6639--6648 (2023). \doi{10.1002/mp.16735}

\bibitem{Ioffe2015}
Ioffe, S., Szegedy, C.: Batch normalization: {Accelerating} deep network training by reducing internal covariate shift. 32nd International Conference on Machine Learning, ICML 2015  \textbf{1} (Feb 2015), \url{http://arxiv.org/abs/1502.03167}

\bibitem{kingmaAdamMethodStochastic2014}
Kingma, D.P., Ba, J.: Adam: {A} {Method} for {Stochastic} {Optimization} (Dec 2014). \doi{10.48550/arXiv.1412.6980}

\bibitem{krizhevskyOneWeirdTrick2014}
Krizhevsky, A.: One weird trick for parallelizing convolutional neural networks (Apr 2014). \doi{10.48550/arXiv.1404.5997}

\bibitem{liuCLIPDrivenUniversalModel2023}
Liu, J., Zhang, Y., Chen, J.N., Xiao, J., Lu, Y., Landman, B.A., Yuan, Y., Yuille, A., Tang, Y., Zhou, Z.: {CLIP}-{Driven} {Universal} {Model} for {Organ} {Segmentation} and {Tumor} {Detection}. In: 2023 {IEEE}/{CVF} {International} {Conference} on {Computer} {Vision} ({ICCV}). pp. 21095--21107 (Oct 2023). \doi{10.1109/ICCV51070.2023.01934}

\bibitem{consortiumMONAIMedicalOpen2024}
{MONAI Consortium}: {MONAI}: {Medical} {Open} {Network} for {AI} (Oct 2024). \doi{10.5281/zenodo.13942962}

\bibitem{mooreExperienceBasedQualityControl2011}
Moore, K.L., Brame, R.S., Low, D.A., Mutic, S.: Experience-{Based} {Quality} {Control} of {Clinical} {Intensity}-{Modulated} {Radiotherapy} {Planning}. International Journal of Radiation Oncology*Biology*Physics  \textbf{81}(2),  545--551 (Oct 2011). \doi{10.1016/j.ijrobp.2010.11.030}

\bibitem{mooreQuantifyingUnnecessaryNormal2015b}
Moore, K.L., Schmidt, R., Moiseenko, V., Olsen, L.A., Tan, J., Xiao, Y., Galvin, J., Pugh, S., Seider, M.J., Dicker, A.P., Bosch, W., Michalski, J., Mutic, S.: Quantifying {Unnecessary} {Normal} {Tissue} {Complication} {Risks} due to {Suboptimal} {Planning}: {A} {Secondary} {Study} of {RTOG} 0126. International Journal of Radiation Oncology*Biology*Physics  \textbf{92}(2),  228--235 (Jun 2015). \doi{10.1016/j.ijrobp.2015.01.046}

\bibitem{nguyenIncorporatingHumanLearned2020}
Nguyen, D., McBeth, R., Sadeghnejad~Barkousaraie, A., Bohara, G., Shen, C., Jia, X., Jiang, S.: Incorporating human and learned domain knowledge into training deep neural networks: {A} differentiable dose-volume histogram and adversarial inspired framework for generating {Pareto} optimal dose distributions in radiation therapy. Medical Physics  \textbf{47}(3),  837--849 (2020). \doi{10.1002/mp.13955}

\bibitem{radfordLearningTransferableVisual2021}
Radford, A., Kim, J.W., Hallacy, C., Ramesh, A., Goh, G., Agarwal, S., Sastry, G., Askell, A., Mishkin, P., Clark, J., Krueger, G., Sutskever, I.: Learning {Transferable} {Visual} {Models} {From} {Natural} {Language} {Supervision} (Feb 2021). \doi{10.48550/arXiv.2103.00020}

\bibitem{shenOperatingTreatmentPlanning2020b}
Shen, C., Nguyen, D., Chen, L., Gonzalez, Y., McBeth, R., Qin, N., Jiang, S.B., Jia, X.: Operating a treatment planning system using a deep-reinforcement learning-based virtual treatment planner for prostate cancer intensity-modulated radiation therapy treatment planning. Medical Physics  \textbf{47}(6),  2329--2336 (2020). \doi{10.1002/mp.14114}

\bibitem{sproutsDevelopmentDeepReinforcement2022}
Sprouts, D., Gao, Y., Wang, C., Jia, X., Shen, C., Chi, Y.: The development of a deep reinforcement learning network for dose-volume-constrained treatment planning in prostate cancer intensity modulated radiotherapy. Biomedical Physics \& Engineering Express  \textbf{8}(4),  045008 (Jun 2022). \doi{10.1088/2057-1976/ac6d82}

\bibitem{talcottBlindedProspectiveStudy2020a}
Talcott, W.J., Lincoln, H., Kelly, J.R., Tressel, L., Wilson, L.D., Decker, R.H., Ford, E., Hartvigson, P.E., Pawlicki, T., Evans, S.B.: A {Blinded}, {Prospective} {Study} of {Error} {Detection} {During} {Physician} {Chart} {Rounds} in {Radiation} {Oncology}. Practical Radiation Oncology  \textbf{10}(5),  312--320 (Sep 2020). \doi{10.1016/j.prro.2020.05.012}

\end{thebibliography}
\end{document}